\documentclass[prb,twocolumn,showpacs,aps,superscriptaddress,floatfix]{revtex4-2}
\usepackage{amsmath}
\usepackage{amssymb}
\usepackage{amsfonts}
\usepackage{bm}
\usepackage[dvips]{graphicx}
\usepackage{color}
\usepackage{ulem}
\usepackage[unicode, pdftex]{hyperref}
\usepackage{verbatim}
\begin{document}
\title{Lifshitz transition in dirty doped topological insulator with the nematic superconductivity}
\author{R.S. Akzyanov}
\affiliation{Dukhov Research Institute of Automatics, Moscow, 127055 Russia}
\affiliation{Moscow Institute of Physics and Technology, Dolgoprudny,
    Moscow Region, 141700 Russia}
\affiliation{Institute for Theoretical and Applied Electrodynamics, Russian
    Academy of Sciences, Moscow, 125412 Russia}

\begin{abstract}
We study the effects of the Lifshitz transition from closed to open Fermi surface in dirty topological insulators with the nematic superconductivity near the critical temperature. We solve linearized Gor'kov equations and find that the nematic superconductor with an open Fermi surface has a lower critical temperature and more susceptible to the disorder than the superconductor with the closed Fermi surface. We propose that correspondence between the critical temperature and stability against the disorder is the general feature of the superconductivity. We investigate the effects of the Lifshitz transition on the competition between superconducting phases in a topological insulator. Open Fermi surface is beneficial for the nematic order parameter $\Delta_4$ in competition with orbital-triplet $\Delta_2$ and disfavors nematic state over the s-wave order parameter. We study Meissner currents in both clean and dirty limits. We found that transition from closed to open Fermi surface increases anisotropy of Meissner currents. Finite disorder suppresses superconducting density stronger than critical temperature. We compare our results with the existing experimental data.    
\end{abstract}
\maketitle
\section{Introduction}
Superconductivity in topological insulators in Bi$_2$Se$_3$ family attracts significant attention due to realization of topological odd-parity superconductivity\cite{Fu2010,Yonezawa2018}. Experiments on Knight shift~\cite{Matano2016}, second critical field~\cite{Pan2016,Kuntsevich2018,Kuntsevich2019}, magnetic torque~\cite{Asaba2017}, show two-fold symmetry of the response that is incommensurate with the crystal symmetry. Such symmetry breaking arises from the nematic superconducting order parameter within $E_u$ representation\cite{Fu2014,Venderbos2016}. Unconventional nematic superconductivity gives rise to several intriguing phenomena such as surface Andreev bound states~\cite{Hsieh2012,Hao2015,Hao2017}, half-quantum vortices~\cite{Zyuzin2017,How2020}, spin (nematic) vortices~\cite{Wu2017}, spontaneous strain and magnetization~\cite{Akzyanov2020a}, vestigial order~\cite{Hecker2018}, unconventional Higgs modes~\cite{Uematsu2019}, anisotropic quasiparticle interference~\cite{Bao2018,Khokhlov2021}.

Anderson theorem~\cite{Anderson1959} does not hold in general for the unconventional superconductivity that results in suppression of the critical temperature $T_c$ by the disorder~\cite{Millis1988}. In Ref.\cite{Smylie2017} it was found that proton irradiation of Nb$_x$Bi$_2$Se$_3$ decreases critical temperature with increasing density of defects. However, only a small part of scattering events contribute to the pair breaking mechanism~\cite{Smylie2017}. Excessive Cu doping of Bi$_2$Se$_3$ brings additional defects into the system that leads to a slight decrease of critical temperature with the increased doping~\cite{Kriener2012,Kawai2020}. 

In Refs.\cite{Nagai2015,Cavanagh2020,Dentelski2020,Sato2020} effects of the disorder on critical temperature of nematic superconductor was studied. It was found that density disorder decreases critical temperature for the nematic order parameter.  These results contradict the results of Ref.~\cite{Andersen2020} where the robustness of the nematic superconductivity against disorder was derived.

Superfluid density is particularly sensitive to the disorder\cite{levitov_book}. In Ref.~\cite{Kriener2012} it was shown that an increase of the disorder suppresses superfluid density in a doped topological insulator. This superfluid density determines the first critical field of the superconductor and London penetration length. In Refs.\cite{Smylie2016,Fang2020} it was shown that the out-of-plane first critical field in Nb$_x$Bi$_2$Se$_3$ is much smaller than the in-plane first critical field. 

The transition from closed Fermi surface to open one is called as Lifshitz transition~\cite{Lifshitz1960}. This transition occurs in underdoped cuprates and has a significant effect on the superconducting properties~\cite{Norman2010,LeBoeuf2011,Perali2012}. Lifshitz transition appears in topological insulator Bi$_2$Se$_3$ upon doping with Nb or Cu~\cite{Lahoud2013,Almoalem2021}. The appearance of this transition coincides with the emergence of superconductivity in the system~\cite{Almoalem2021}. 

In our paper, we answer the question of how Lifshitz transition affects critical temperature, stability against the disorder, and Meissner currents in a topological insulator with the nematic superconductivity.
This paper is organized as follows. In Sec.~\ref{lifshitz_sec_model} we introduce Hamiltonian of topological insulator with the nematic superconductivity and introduce a model for a Lifshitz transition. In Sec.~\ref{lifshitz_sec_clean} we solve linearized Gor'kov equations for Green's functions in a clean limit and calculate the critical temperature in case of closed and open Fermi surfaces of the normal state. In Sec.~\ref{sec_lifshitz_selfenergy} we calculate self-energy that arises due to scattering from the randomly distributed scalar disorder and analyze the effects of the disorder on the critical temperature. In Sec.~\ref{lifshitz_sec_spectral} we discuss general properties of the robustness of the superconducting states against the disorder. We establish a general connection between critical temperature and stability against the disorder and tie it with the conception of the superconducting fitness. In Sec.~\ref{sec_meissner} we calculate superfluid density in clean and dirty limits and show how Lifshitz transition affects the anisotropy of the first critical field. In the Discussion, we compare our results with the experimental results and other works within the field.
\section{Model}\label{lifshitz_sec_model}

We give a short summary of derivation of the Hamiltonian of the normal state of topological insulators from Refs~\cite{Zhang2009, Liu2010}. Crystal structure of Bi$_2$Se$_3$ consists of layers of Bi and Se. Five such layer form a one quintuple layer. These quintuple layers interact weakly through van der Waals forces. Thus, interactions within quintuple layer are the strongest one. In each quintuple layer central layer consist of Se atoms that is sandwiched by Bi layers and Se layers are outermost. Outermost orbitals of Bi ($6s^2 6p^3$) and Se ($4s^2 4p^4$) are p orbitals and we can neglect other orbitals. Hybridization between Bi and Se orbitals leads to the formation of new hybridized orbitals  of  bismuth, Bi and Bi$^*$, selenide Se, Se$^*$ and Se0.  Due to presence of the inversion crystal symmetry it is convenient to consider bonding and anti-bonding states with the definite parity. State $P1^{\pm}=$(Bi $\pm$ Bi$^*$)/$\sqrt{2}$ corresponds to the bonding $+$ or anti-bonding $-$ state of Bi orbitals, state $P2^{\pm}=$(Se $\pm$ Se$^*$)/$\sqrt{2}$ corresponds to the bonding $+$ or anti-bonding $-$ state of Se orbitals. Here $\pm$ corresponds to the parity of the state. After taking into the account of hybridization between Bi(Se) and Bi$^*$(Se$^*$) orbitals it is found that bonding state of $P1^+$ and anti-bonding state of $P2^-$ are closest to the Fermi level. The crystal has a layered structure along z direction which is different from the x or y directions. Thus, crystal field leads to the energy splitting between $p_z$ and $p_x,p_y$ orbitals. It is found that $p_z$ orbitals form conduction  $P1^+_{p_z}$ and valence bands $P2^-_{p_z}$ prior to consideration of the spin-orbit interaction. Strong spin-orbit interaction pushes energy of $P1^+_{p_z}$ down and $P2^-_{p_z}$ up. At some value of spin-orbit interaction orbitals with the opposite parity crosses that leads to the band inversion. This band inversion is a signature of the topological insulator. This transition occurs at the time-reversal invariant symmetric $\Gamma$ point that is located at the center of the Brillouen zone $\Gamma(0,0,0)$ (here (a,b,c) show location of the point with the respect to the primitive lattice vectors). At this points orbitals are closest to the Fermi level. Now, low-energy effective Hamiltonian $H_N(\mathbf{k})$ can be obtained by fitting kp expansion in $(P1^+_{p_z},P2^-_{p_z})$ basis near the $\Gamma$ point to the DFT calculations~\cite{Zhang2009, Liu2010}
\begin{equation}
 H_N(\mathbf{k})\!=\! -\mu \!+\! m\sigma_z\!+\!v\sigma_x(s_xk_y\!-\!s_yk_x)\!+\!v_zk_z\sigma_y.
\end{equation}

Here Pauli matrices $s_x$, $s_y$, $s_z$ act in the spin space $(\uparrow,\downarrow)$, Pauli matrices $\sigma_x$, $\sigma_y$, $\sigma_z$ act in the space of inverted orbitals of Bi and Se atoms near the Fermi level $(P1^+_{p_z},P2^-_{p_z})$, $\mu$ is the chemical potential, $2m$ is the value of the single-electron gap at half-filling $\mu=0$ at $\Gamma$ point ($\bf{k}=0$), $v$ is the in-plane Fermi velocity within the main $(\Gamma M, \Gamma K)$ plane that is parallel to the plane of Bi and Se layers, $v_z$ is the Fermi velocity along $\Gamma Z$ direction that is perpendicular to the orientation of Bi and Se layers. Away from the $\Gamma$ point new terms in the Hamiltonian arise that leads to $k$ dependence of the parameters of the Hamiltonian and emergence of the hexagonal warping which will be discussed in Sec.~\ref{lifshitz_sec_spectral}. It is worth to mention that linear dispersion along $z$ direction works well even away from $\Gamma$ point, see Ref.\cite{Liu2010} 

Spectrum of the normal state is given by
\begin{eqnarray}
E({\bf k})=-\mu \pm \sqrt{m^2+v^2k_x^2+v^2k_y^2+v_z^2k_z^2},
\end{eqnarray}
 Closed Fermi surface forms an ellipsoid that is elongated along z direction since $v_z<v$. This ellipsoid can be parametrized by an ellipsoid coordinates $(vk_x,vk_y,v_zk_z)=\sqrt{\mu^2-m^2}(\cos \varphi \sin \theta ,\sin \varphi \sin \theta, \cos \theta)$. In case of closed Fermi $\varphi \in (0,2\pi)$ and $\theta \in (0,\pi)$.

\begin{figure}[ht]\label{lifhitz_fs}
\includegraphics[width=0.35\linewidth]{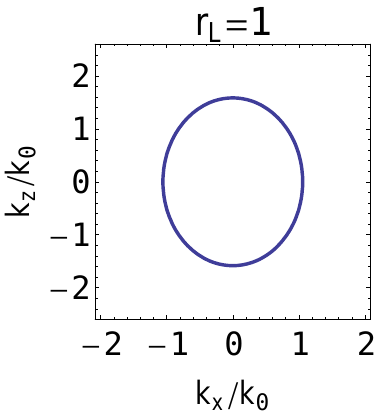}
\includegraphics[width=0.3\linewidth]{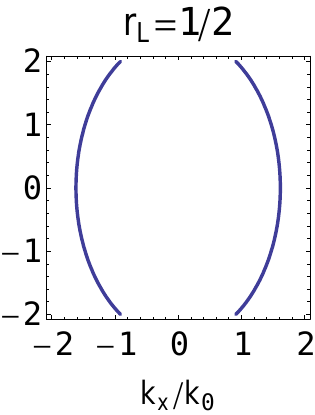}
\includegraphics[width=0.3\linewidth]{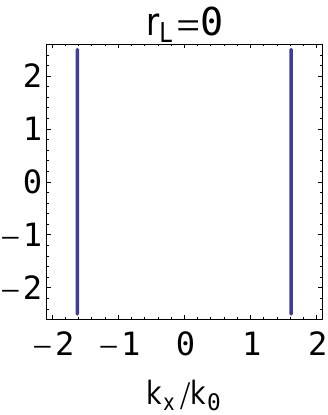}
\caption{Fermi surface in $(k_x/k_0,k_z/k_0)$ plane for $k_y=0$ for different values of Lifshitz parameter $r_L$. Left figure corresponds to the closed Fermi surface $r_L=1$ with $\mu=1.5m$, central figure to corrugated cylinder $r_L=1/2$ with $\mu=2.8m$, right figure to cylindrical Fermi surface $r_L=0$ with $\mu=2.8m$ and $v_z=0$. Here $k_0=m/v$, upper and lower boundaries in $k_z$ direction shows the the boundary for the first Brillouin zone. Boundary for the first Brillouin zone in $k_x$ direction is $c/a \sim 7$ (here a and c are lattice constants) times larger than boundary in $k_z$ direction and is not shown here. }
\label{lifshitz_fs}
\end{figure}

Lifshitz transition from closed to open Fermi surface occurs if size of Brillouen zone $k_c$ becomes smaller than Fermi momentum $v_zk_c<\sqrt{\mu^2-m^2}$. In ellipsoid coordinates this results that angle $\theta \in (\theta_L,\pi-\theta_L)$ where $\cos \theta_L=\textrm{min}\,(1,v_zk_c/\sqrt{\mu^2-m^2})$. Here we introduce parameter $r_L=\cos \theta_L$ that controls Lifshitz transition. If $r_L=1$ then Fermi surface is closed. In case of $r_L<1$ Fermi surface becomes open. Case $r_L=0$ corresponds to the purely cylindrical Fermi surface. Fermi surface $E(\mathbf{k})=0$ is shown at Fig.~\ref{lifshitz_fs} for different values of $r_L$. Note, that obtained Fermi surface is similar to the experimental and DFT calculated Fermi surfaces~\cite{Lahoud2013,Almoalem2021}.

In our work, we suppose that electron-phonon interaction is short-range and has no dependence on the Lifshitz transition. In Ref.~\cite{Wang2019} it was shown that near the Lifshitz transition electron-phonon coupling is enhanced along [001] direction. Also, electron-phonon coupling is singular along $k_z$ direction and isotropic in $(k_x,k_y)$ plane. It means that electrons with the small $k_z$ momentum have strongest coupling. Anisotropic singular coupling $g=g(k_z)$ can be modeled by step function with the size $k_c$. This anisotropic coupling results in the same effects on the superconductivity as a Lifshitz transition except for the density of states $\rho(\mu)$ is unchanged.

Form of the Fermi surface has a significant impact on the density of states. We suppose that only the states near the Fermi surface contribute and calculate density of states at zero frequency as 

\begin{equation}
    \rho(\mu)= -\frac{1}{\pi}\textrm{Im}\,\textrm{Tr}\sum_k\,G_0(\omega\rightarrow +0)=\frac{r_L\mu\sqrt{\mu^2-m^2}}{v^2v_z\pi^2},
\end{equation}
where Green's function of the normal state $G_{0}=[i\omega- H_N(\mathbf{k})]^{-1}$ is 
\begin{eqnarray}\label{eq_bare_green}
G_{0}=\frac{-i\omega-\mu-m\sigma_z-\!v\sigma_x(s_xk_y\!-\!s_yk_x)\!-\!v_zk_z\sigma_y}{m^2+v_z^2k_z^2+v^2(k_x^2+k_y^2)-(\mu+i\omega)^2}.
\end{eqnarray}
We plot the density of states as a function of doping $\mu$ in the absence and presence of the Lifshitz transition. Increase of the chemical potential increase density of states by the square law $\rho \propto \mu^2$ for $\mu \gg m$ in case of closed Fermi surface $r_L=1$. After the Lifshitz transition $r_L<1$ density of states increases linearly with the increase of chemical potential $\rho \propto \mu$. This picture is in qualitative agreement with the density functional theory calculations: fast growth of the density of states before Lifshitz transition and slow growth after it.

\begin{figure}[ht]
\includegraphics[width=1\linewidth]{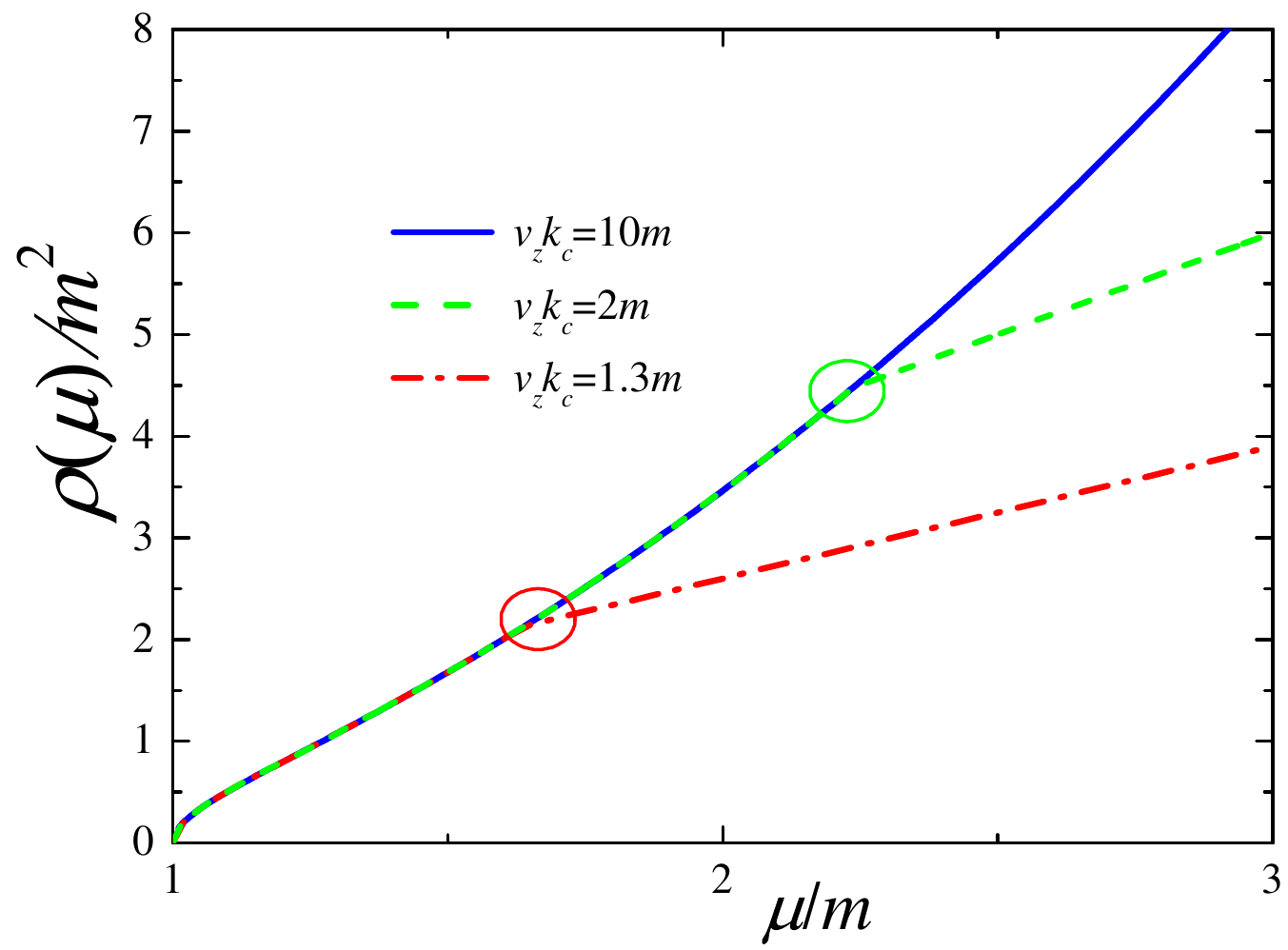}
\caption{Density of states $\rho(\mu)$ as a function of chemical potential $\mu$ for different values of the Brillouen zone size $v_zk_c$. Red and green circle shows point of the Lifshitz transition for $v_zk_c=1.3$ and $v_zk_c=2$ respectively. 
}
\label{lifshitz_rho}
\end{figure}

We consider the system with the odd-parity order parameter within $E_u$ representation. Such nematic order parameter couples electrons from the different orbitals with the same spin and preserves time-reversal symmetry. This order parameter has a vector structure $\Delta=(\Delta_x,\Delta_y)$ that differs from other possible pairings that transforms as a scalar under rotation. In Nambu basis $(\Psi(k),is_y\Psi^{*}(-k))$ doped topological insulator with the superconductivity can be described by Hamiltonian (see Ref.~\cite{Yip2013} for discussion about form of the order parameters in different basis) 
\begin{eqnarray}\label{BdG}
H_{\textrm{BdG}}(\mathbf{k})=H_N(\mathbf{k})\tau_z+\hat{\Delta}\tau_x, \\
\hat{\Delta}=(\Delta_xs_x+\Delta_ys_y)\sigma_y,
\end{eqnarray}
where Pauli matrices $\tau_i$ act in a Nambu particle-hole space. 

\section{Gor'kov equations for the clean case}\label{lifshitz_sec_clean}
We start with the Gor'kov equations in general case in Nambu basis $\Psi=(\Psi(k),is_y\Psi^{*}(-k))$.
Green's functions can be obtained by solving Gor'kov equations
\begin{eqnarray}
(i\omega-\hat{H}_{\textrm{BdG}})\hat{G}_0=\hat{1},
\end{eqnarray}
where Hamiltonian in Nambu space expresses as
\begin{eqnarray}
i\omega-\hat{H}_{\textrm{BdG}}=
\begin{pmatrix}
i\omega-H_N(k) & -\hat{\Delta} \\
-\hat{\Delta}^{\dagger} & i\omega+s_y H_N^{*}(-k)s_y 
\end{pmatrix}, 
\end{eqnarray}
and Green's function $\hat{G}_0$ as
\begin{eqnarray}
\hat{G}_0=
\begin{pmatrix}
G_{0e} & F_0 \\
\bar{F_0} & G_{0h}
\end{pmatrix}.
\end{eqnarray}
Here $G_{0e} (G_{0h})$ is the normal part of electrons (holes) and $F_0$ is the anomalous part of the Green's functions.
Hamiltonian of the normal state is $H_N(k)$, $\hat{\Delta}$ is the superconducting order parameter, $\omega=\pi T(2n+1)$ is the fermionic Matsubara frequency for temperature $T$, $\hat{1}$ is identity matrix. For the anomalous Green's function that is responsible for the superconducting correlations in the system we have 
\begin{eqnarray}
F_0=[i\omega-H_N(k)]^{-1}\hat{\Delta}G_{0h}, \\
\bar{F_0}=[i\omega+s_y H_N^{*}(-k)s_y]^{-1}\hat{\Delta}^{\dagger}G_{0e}.
\end{eqnarray}
Normal part of Green's function expresses as
\begin{eqnarray}
G_{0e}=(1-G_0\hat{\Delta} G_0^{*}\hat{\Delta}^{\dagger})^{-1}G_0, \\
G_{0h}=(1-\bar{G}_0\hat{\Delta}^{\dagger} G_0\hat{\Delta})^{-1}\bar{G}_0,
\end{eqnarray}
where we introduce bare Green's functions of the normal state as
\begin{eqnarray}
G_0=[i\omega-H_N(k)]^{-1}, \\
\bar{G}_0=[i\omega+s_y H_N^{*}(-k)s_y]^{-1}.
\end{eqnarray}

Near critical temperature $T_c$ we can keep only linear in order parameter $\hat{\Delta}$ terms in the Green's functions. We consider a system with the time-reversal symmetry $s_y H_N^{*}(-k)s_y=H_N(k)$ that results in 
\begin{eqnarray}
G_{0e}=G_0, \quad G_{0h}=\bar{G}_0, \\
F_0=G_0\hat{\Delta}\bar{G}_0,\quad \bar{F_0}=\bar{G}_0\hat{\Delta}^{\dagger}G_0.
\end{eqnarray}
Note, that $\bar{G}_0(\omega)=-G_0(-\omega)$.  
We will use these linearized in $\hat{\Delta}$ expression for our calculations.
 In case of topological insulator we use Eq.~\ref{eq_bare_green} for $G_0$. Linearized anomalous Green's functions are written as
\begin{eqnarray}
F_0=\frac{2f(k,\Delta,\omega)}{B_N} \nonumber\\
\bar{F_0}=-2f(k,\Delta^{\dagger},-\omega)/B_N 
\end{eqnarray}
where 
\begin{widetext}
\begin{eqnarray}\label{anomalous_green_function}
f(k,\Delta,\omega)=-v_zk_z\mu(\Delta_xs_x+\Delta_ys_y) + mv(k_x\Delta_x+k_y\Delta_y)s_z+vv_zk_z(k_x\Delta_y-k_y\Delta_x)\sigma_x-m\omega(\Delta_xs_x+\Delta_ys_y)\sigma_x+\nonumber\\ \nonumber
J_xs_x\sigma_y+J_ys_y\sigma_y+v\omega(k_y\Delta_x-k_x\Delta_y)\sigma_z-v_zk_zm(\Delta_xs_x+\Delta_ys_y)\sigma_z+v\mu(k_x\Delta_x+k_y\Delta_y)s_z\sigma_z,
\\\nonumber
J_x=(\Delta_x(m^2-\mu^2-\omega^2+v^2(k_y^2-k_x^2)-v_z^2k_z^2)-2v^2k_xk_y\Delta_y)/2,
\\\nonumber
J_y=(\Delta_y(m^2-\mu^2-\omega^2+v^2(k_x^2-k_y^2)-v_z^2k_z^2)-2v^2k_xk_y\Delta_x)/2, \\
B_N=(m^2+v_z^2k_z^2+v^2(k_x^2+k_y^2)-\mu^2+\omega^2)^2+4\mu^2\omega^2. \quad
\end{eqnarray}
\end{widetext}

\begin{figure}[ht]
\includegraphics[width=1\linewidth]{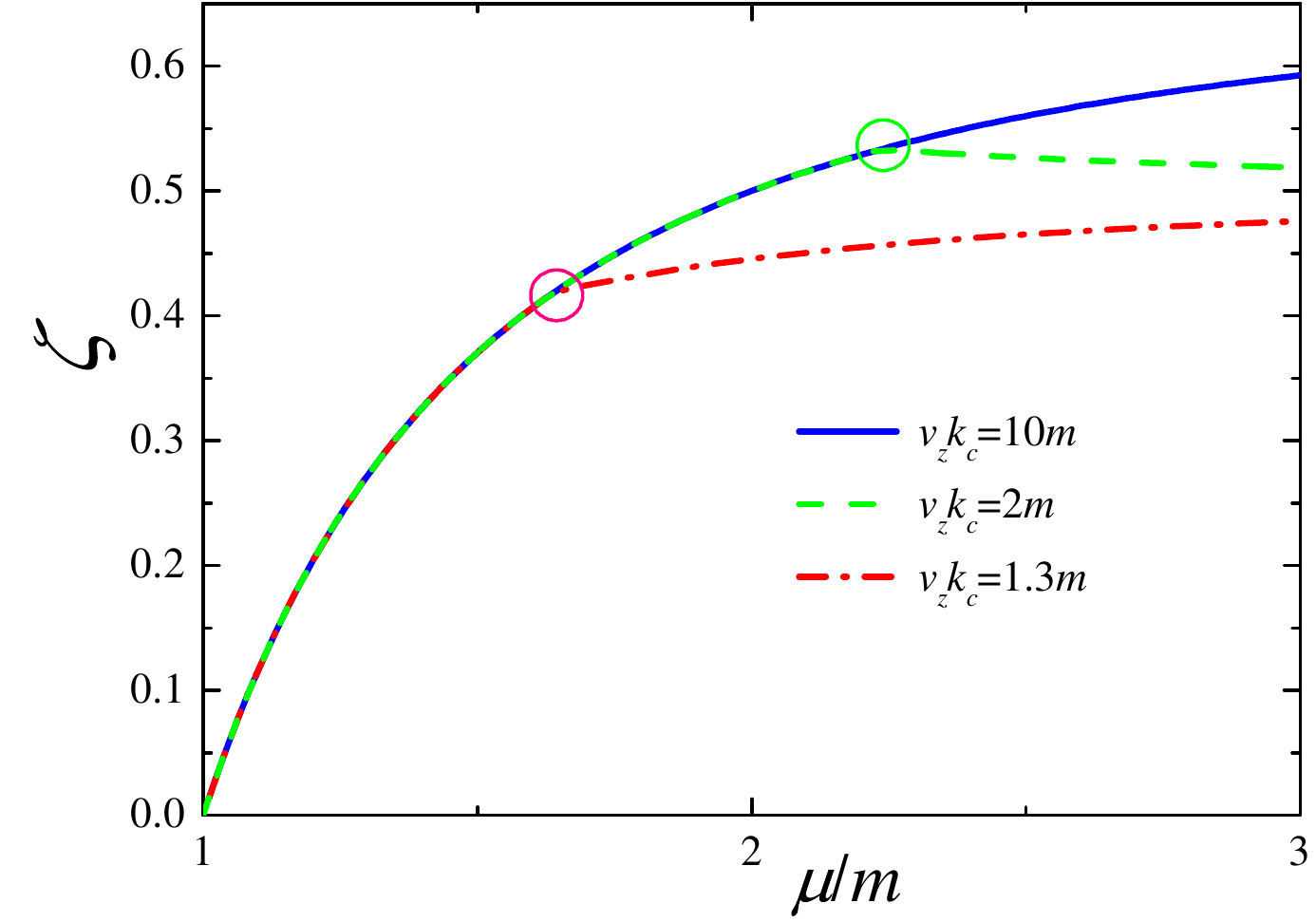}
\caption{Parameter $\zeta$ as a function of chemical potential $\mu$ for different values of the Brillouen zone size $v_zk_c$. Red and green circle shows point of the Lifshitz transition for $v_zk_c=1.3$ and $v_zk_c=2$ respectively. }
\label{lifshitz_zeta}
\end{figure}

Anomalous Green's function looks quite complex. However, only $\tilde{f}(k,\Delta,\omega)$ part of the anomalous Green's function contribute to the integral over the Brillouen zone
\begin{eqnarray}
\tilde{f}(k,\Delta,\omega)=-m\omega(\Delta_xs_x+\Delta_ys_y)\sigma_x+ \nonumber \\
(\Delta_xs_x\sigma_y+\Delta_ys_y\sigma_y)(m^2-\mu^2-\omega^2-v_z^2k_z^2)/2.
\end{eqnarray}
Self-consistent equation for the nematic order parameter $\Delta_i=-gT/4 \sum_{\omega,k} \textrm{Tr}\,[\sigma_ys_i F_0]$, $i=x,y$ is written as
\begin{eqnarray}
\Delta_{x(y)}=-gT\sum\limits_{\omega}\int \frac{d^3k}{(2\pi)^3} \Delta_{x(y)} \times \\ \nonumber \frac{(m^2-\mu^2-\omega^2-v_z^2k_z^2)}{(m^2+v_z^2k_z^2+v^2(k_x^2+k_y^2)-\mu^2+\omega^2)^2+4\mu^2\omega^2},
\end{eqnarray}
where $g$ is the coupling strength.
Different orientations of the nematicity (or even any superposition of $\Delta_{x}$ and $\Delta_{y}$) have the same $T_c$. So, we can consider only one orientation of the nematicity $\Delta_x$ without loss of the generality. We can see from this expression that momentum along $z$ direction has a distinct impact on the value of the critical temperature.
Integration over the momentum gives us
\begin{eqnarray}\label{clean_tc}
\Delta= \frac{\pi gT \rho(\mu)}{4}\zeta\sum\limits_{\omega} \frac{\Delta}{|\omega|}, 
\end{eqnarray}
where we introduce parameter $\zeta$ as
\begin{equation}\label{zeta_parameter}
    \zeta=\frac{(1+r_L^2/3)(\mu^2-m^2)}{ 2\mu^2}.
\end{equation}
In a weak coupling approximation critical temperature expresses as $T_{c0} \simeq 1.14 \omega_D \exp[-4/g\zeta \rho(\mu)]$, where $\omega_D$ is Debye cut-off. We see that this expression is identical to the expression for the critical temperature of the s-wave superconductor with the renormalized by $\zeta$ coupling strength. We plot parameter $\zeta$ as a function of the chemical potential for different values of the Brillouin zone cutoff $k_c$ at Fig.~\ref{lifshitz_zeta}. We see that after the Lifshitz transition parameter $\zeta$ has slow growth in comparison with the case of closed Fermi surface and can even decrease with the increase of the chemical potential. In case of closed Fermi surface $r_L=1$ we have $\zeta=2/3(1-m^2/\mu^2)$ while cylindrical one $r_L=0$ gives us smaller value $\zeta=1/2(1-m^2/\mu^2)$.  
\section{Effects of scalar impurities}\label{sec_lifshitz_selfenergy}
In this section we study effects of a random charged impurities. We will describe disorder by a potential $V_{\textrm{imp}}=u_0\tau_z \sum_i \delta(\mathbf{r}-\mathbf{R}_j)$, where $\delta(\mathbf{r})$ is the Dirac delta function, $\mathbf{R}_j$ are the positions of the randomly distributed point-like impurities with the local potential $u_0$ and concentration $n_i$, $\tau_z$ shows that electrostatic potential acts contrary on electrons and holes. We assume that the disorder is Gaussian, that is, $\langle V_{\textrm{imp}} \rangle=0$ and $\langle V_{\textrm{imp}}(\mathbf{r}_1) V_{imp}(\mathbf{r}_2) \rangle=n_i u_0^2 \delta (\mathbf{r}_1-\mathbf{r}_2)$. 

Self-energy is calculated as
\begin{equation}\label{hatsigma}
\hat{\Sigma}=n_iu_0^2 \sum\limits_k\tau_z\hat{G}\tau_z, \\
\end{equation}
and has following matrix structure
\begin{eqnarray}
\hat{\Sigma}=
\begin{pmatrix}
\Sigma_{e} & \Sigma_F \\
\bar{\Sigma}_F &\Sigma_{h}
\end{pmatrix}.
\end{eqnarray}
We calculate self-energy of the normal state in a first Born approximation as
\begin{eqnarray}
\Sigma_{e(h)}=n_iu_0^2 \sum\limits_kG_{0e(h)}.
\end{eqnarray}
 Self-energy of the normal part has two components due to strong hybridization between orbitals~\cite{Akzyanov2020c}. We assume that Debye cut-off is small $\omega_D \ll \sqrt{\mu^2-m^2}$ and calculate self-energies at infinitesimally small frequency $\Sigma_e(\omega)=\Sigma_e(\omega \rightarrow 0)$ that means that we keep only imaginary part of the self-energy. Real part of the self-energy leads to the small addition to $\mu$ and $m$ which we can neglect. Under this assumption self-energy of the normal state is
\begin{eqnarray}
\Sigma_{e(h)}(\omega)=\Sigma_{e(h)0}+\Sigma_{e(h)m} \sigma_z,\\
\Sigma_{e(h)0}(\omega)=-i\Gamma_0, \quad \Sigma_{e(h)m}(\omega)=-i\Gamma_z, \\
\Gamma_0=\textrm{sgn}\,(\omega)\, n_iu_0^2 \frac{\pi\rho(\mu)}{4}, \quad \Gamma_z=\Gamma_0\frac{m}{\mu}. 
\end{eqnarray}

Disorder averaged Green's function of the normal state $G_{e(h)}$ can be found using Dyson equation $G_{e(h)}^{-1}=G_{0e(h)}^{-1}-\Sigma_{e(h)}$. This results in renormalization of the Matsubara frequency $\omega \rightarrow \omega + \Gamma_0$ and single-electron gap $m \rightarrow m-i\Gamma_z$. 

We calculate anomalous self-energy using disorder-averaged Green's functions of the normal state as
\begin{eqnarray}\label{sigmaf_equation}
\Sigma_F=-n_iu_0^2 \sum\limits_kG_{e}\hat{\Delta}G_h.
\end{eqnarray}
Here sign $-$ appears due to $\tau_z$ factor in the impurity potential that appears due to different charge of electrons and holes.
Anomalous self-energy has two components
\begin{eqnarray}
\Sigma_F=(\Delta_xs_x+\Delta_ys_y)(\sigma_y\Sigma_{F1}+\sigma_x\Sigma_{F2}), \nonumber\\
\Sigma_{F1}(\omega)=\bar{\Gamma}\zeta/\omega, \quad \Sigma_{F2}(\omega)=\bar{\Gamma}\frac{m}{\mu^2}. 
\end{eqnarray}
Here $\bar{\Gamma}=\Gamma_0(1+m^2/\mu^2)$ is the effective scattering rate and parameter $\zeta$ is defined by Eq.~\ref{zeta_parameter}.
As we can see, $\Sigma_{F1}$ renormalizes value of the order parameter $\hat{\Delta}$, while $\Sigma_{F2}$ brings new term $i\sigma_z\hat{\Delta}$. In general, this means that ground state is a mixture between spontaneously generated order parameter $\hat{\Delta}$ and disorder induced term $i\sigma_z\hat{\Delta}$. However, this new term is small since $\Sigma_{F1} \propto 1/|\omega|$ and $\Sigma_{F2} \propto m/\mu^2$. Thus, in case of $\omega_D/\mu \ll 1$ ground state has only component $\hat{\Delta}$ in the order parameter.

Disorder-averaged anomalous Green's functions is calculated as $F=G_e(\hat{\Delta}+\Sigma_F)G_h$. Self-consistent equation $\Delta_i=-gT/4 \sum_{\omega,k} \textrm{Tr}\,[\sigma_ys_i F]$ leads to
\begin{widetext}
\begin{eqnarray}
\Delta_{x(y)}\!=\!-gT\sum\limits_{\omega}\!\int\! \frac{d^3k}{(2\pi)^3} 
\frac{\Delta_{x(y)}[1+\Sigma_{F1}(\omega)][m^2-\mu^2-(\omega+\Gamma_0)^2+\Gamma_z^2-v_z^2k_z^2]+2\Sigma_{F2}(\omega)(m\omega-\Gamma_z\mu+m\Gamma_0)}{[m^2+v_z^2k_z^2+v^2(k_x^2+k_y^2)-\mu^2+(\omega+\Gamma_0)^2-\Gamma_z^2]^2+4[\mu(\omega+\Gamma_0)+m\Gamma_z]^2}.
\end{eqnarray}
\end{widetext}
For weak scattering $|\Gamma_0|,|\Gamma_z| \ll \mu$ and $\omega_D \ll \mu$ we calculate to
\begin{equation}
\Delta=\frac{\pi g\zeta\rho(\mu)T}{4} \sum\limits_{\omega} \frac{\tilde{\Delta}}{|\tilde{\omega}|},
\end{equation}
where renormalized by the disorder Matsubara frequency $\tilde{\omega}$ and order parameter $\tilde{\Delta}$ are
\begin{equation}
\tilde{\omega}=\omega+\bar{\Gamma},\quad
\tilde{\Delta}=\Delta\left(1+\frac{\bar{\Gamma}\zeta}{\omega}\right).
\end{equation}

If we substitute $\tilde{\omega}$ and $\tilde{\Delta}$ back into the equation for the anomalous Green's function $F$ we arrive to the different $\tilde{\Delta}$. Self-consistent procedure leads for $\tilde{\Delta}$ in following equations
\begin{eqnarray}
\tilde{\omega}=\omega+\bar{\Gamma}, \quad\tilde{\Delta}=\Delta+\tilde{\Delta}\bar{\Gamma}\zeta/\tilde{\omega},
\end{eqnarray}
or
\begin{eqnarray}\label{disorder_renormalized_delta}
\tilde{\omega}=\omega+\bar{\Gamma}, \quad
\tilde{\Delta}=\Delta/\left(1-\bar{\Gamma} \zeta/\tilde{\omega}\right).
\end{eqnarray}
Self-consistent equation in a self-consistent approximation is written as
\begin{eqnarray}\label{disorder_tc}
\Delta= \frac{\pi g \rho(\mu) \zeta T}{4}\sum\limits_{\omega} \frac{\Delta}{|\omega+(1-\zeta)\bar{\Gamma}|}.
\end{eqnarray}
This equation leads to the Abrikosov-Gor'kov equation for a critical temperature
\begin{equation}
\ln \frac{T_c}{T_{c0}}=\Psi(1/2)-\Psi\left(1/2+\frac{\bar{\Gamma}(1-\zeta)}{2\pi T_c}\right).    
\end{equation}
where $\Psi(x)$ is the digamma function. Critical temperature is completely suppressed at
\begin{equation}
(1-\zeta)\bar{\Gamma}_{c}= 0.88 T_{c0}.
\end{equation}

Nematic superconductivity is suppressed by the large disorder that confirms results of Refs.\cite{Cavanagh2020,Dentelski2020,Sato2020}. The critical temperature depends on the parameter $\zeta$ that determines both the critical temperature in a clean case and robustness against the disorder according to Eqs.~\ref{clean_tc} and~\ref{disorder_tc}. This parameter depends on the shape of the Fermi surface. Closed Fermi surface $r_L=1$ gives $\zeta=2/3$ for $\mu \gg m$ that is consistent with the results of Ref.\cite{Cavanagh2020}. Cylindrical Fermi surface $r_L = 0$ gives $\zeta=1/2$. It means that a closed Fermi surface is more robust against the disorder and has a higher critical temperature for the same density of states than a cylindrical one.  

\begin{figure}[ht]
\includegraphics[width=1\linewidth]{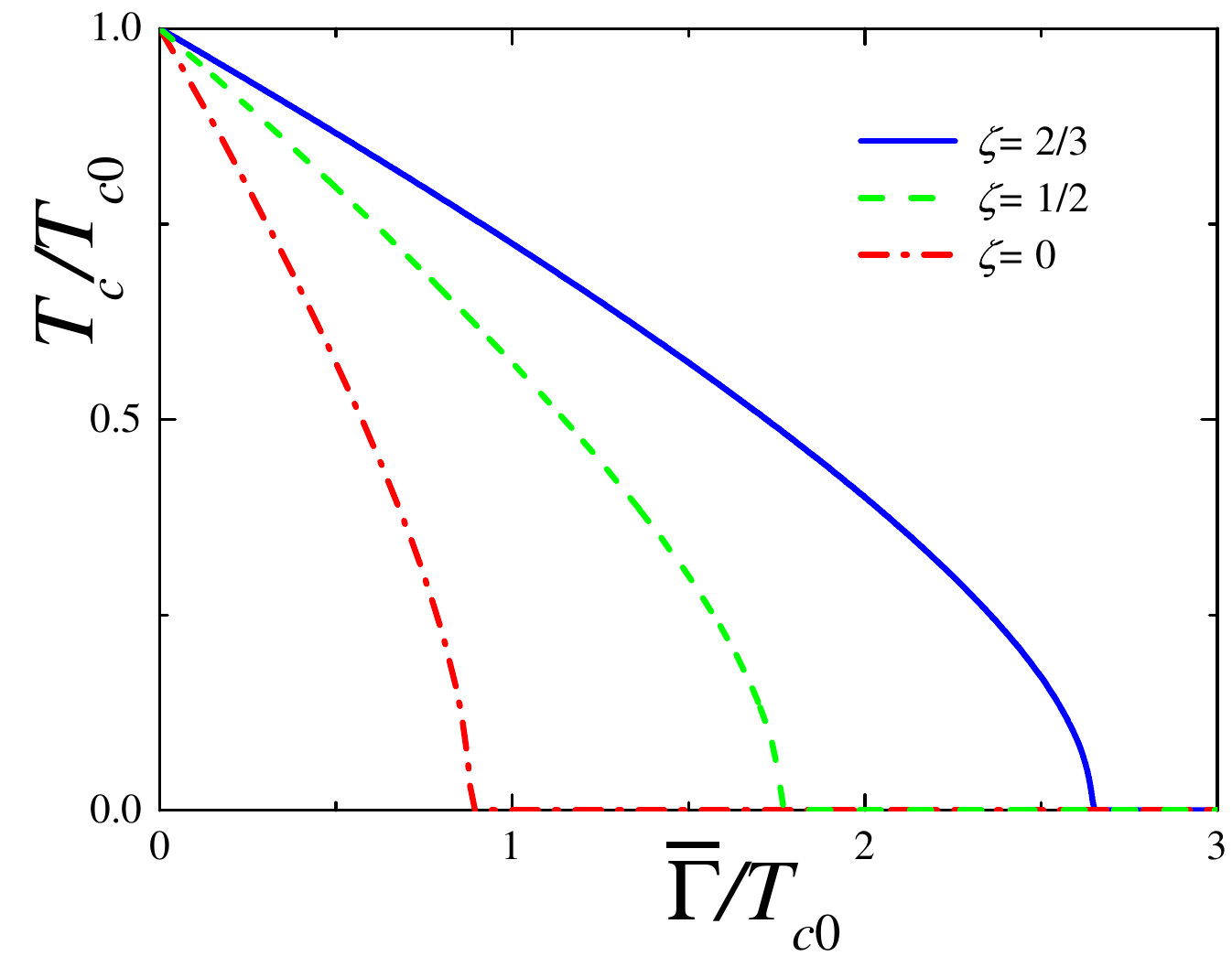}
\caption{Critical temperature $T_c$ as a function of disorder $\bar{\Gamma}$ for different values of parameter $\zeta$.  }
\label{lifshitz_tcg}
\end{figure}

\section{Spectral representation}\label{lifshitz_sec_spectral}
In order to get insides how parameter $\zeta$ ties together both critical temperature and robustness against the disorder we write down Gor'kov equations in a spectral representation. We suppose that matrix $\hat{A}$ determine spin and orbital structure of the order parameter $\hat{\Delta}=\Delta \hat{A}$ where $\Delta$ is the scalar that determine value of the order parameter and $\hat{A}^{\dagger}\hat{A}=1$. A is the $n \times n$ matrix where n is the number of bands that contribute to the order parameter. We consider the case when only single band of the normal state $\epsilon_l$ crosses Fermi level $\mu$. We consider that this band is degenerate $p_l$ times $H_N \psi_i =\epsilon_l \psi_i$, $i\in {\bf p_l}=1,..,p_l$ where ${\bf p_l}$ denotes set of eigenvectors with energy $\epsilon_l$. We also assume presence of the time-reversal symmetry.
Green's function in a normal state and anomalous Green's function are given by~\cite{Cavanagh2020}
\begin{eqnarray}\label{spectral_fo}
G_{e0}=\sum\limits_j\frac{P_j}{i\omega+\mu-\epsilon_j},\quad
G_{h0}=\sum\limits_j\frac{P_j}{i\omega-\mu+\epsilon_j},\\
F_0= \Delta\sum\limits_{i,j}\frac{P_i \hat{A} P_j}{(i\omega+\mu-\epsilon_i)(i\omega-\mu+\epsilon_j)},
\end{eqnarray}
where projector on the band with energy $\epsilon_i$ is given by $P_i= |\psi_i\rangle \langle \psi_i|$. Using assumption that only level with the energy $\epsilon_l$ crosses Fermi level, self-consistent equation for the value of the order parameter $\Delta$ in a clean limit is written as
\begin{eqnarray}
\Delta=-\frac{gT}{n}\sum\limits_{\omega,k} \textrm{Tr} [\hat{A}^{\dagger} F_0]=\frac{\pi g \rho(\mu)}{n} \zeta \sum_{\omega}\frac{\Delta}{|\omega|}.
\end{eqnarray}
Here density of states $\rho(\mu)= \omega p_l/\pi\sum_k 1/(\omega^2+(\epsilon_l-\mu)^2)|_{\omega \rightarrow +0}$. Parameter $\zeta$ expresses through Fermi surface projected order parameter $A_p$ as
\begin{eqnarray}\label{proj_order_parameter}
\zeta=\textrm{Tr}\langle A_p^{\dagger} A_p\rangle_{FS}/p_l,\\
A_{pij}=\langle \psi_i |\hat{A}| \psi_j \rangle,
\end{eqnarray}
where $A_{pij}$ is calculated for the states with eigenenergy that crosses Fermi level $i,j \in {\bf p_l}$. Similar expression can be obtained through direct calculation of the Cooper susceptibility for general case of momentum dependent order parameter $A=A_k$~\cite{Fu2009}.

If we neglect scattering between the different states within the band then self-energy of the normal state in presence of scalar disorder is diagonal $\Sigma_N=-i\Gamma_0 \hat{1}$ where $\Gamma_0=\textrm{sgn}\,(\omega)n_iu_0^2 \pi\rho/n$. Dyson equation for the normal state $G_N^{-1}=G_{N0}^{-1}-\Sigma_N$ shows that disorder renormalizes Matsubara frequency $\omega\rightarrow \tilde{\omega}=\omega+\Gamma_0$.

Leading contribution of the anomalous self-energy $\Sigma_F=-n_iu_0^2 \sum_k F_0(\tilde{\omega})$ that renormalizes value of the order parameter is
\begin{eqnarray}
\Sigma_F=\sigma_F \hat{A}^{\dagger}, \\
\sigma_F=-n_iu_0^2\Delta \sum\limits_k \textrm{Tr} [\hat{A}^{\dagger} F_0]/n = \Delta \zeta \Gamma_0/\omega.
\end{eqnarray}
From the Dyson equation $G^{-1}=G_0^{-1}-\Sigma$ we can see that in presence of the disorder anomalous Green's function given by Eq.~\ref{spectral_fo} can be obtained by the substitution $\hat{\Delta} \rightarrow \hat{\Delta}+\Sigma_F $. As a result, self-consistent equation is
\begin{eqnarray}
\Delta=\pi g \zeta\rho/n\sum_{\omega}\frac{\tilde{\Delta}}{|\tilde{\omega}|},
\end{eqnarray}
where
\begin{eqnarray}
\tilde{\omega}=\omega+\Gamma_0,\quad
\tilde{\Delta}=\Delta+\Delta \zeta \Gamma_0/\omega.
\end{eqnarray}
Self-consistent procedure leads to
\begin{eqnarray}
\tilde{\omega}=\omega+\Gamma_0,\quad
\tilde{\Delta}=\Delta+\tilde{\Delta} \zeta \Gamma_0/\omega,
\end{eqnarray}
and we arrived to the expression
\begin{eqnarray}
\Delta\!=\!-\frac{gT}{n}\sum\limits_{\omega,k}\! \textrm{Tr} [\hat{A}^{\dagger} F]\!=\!\frac{\pi g \zeta\rho}{n}\!\sum\limits_{\omega}\!\frac{\Delta}{|\omega+(1-\zeta)\Gamma_0|},
\end{eqnarray}
which is similar to Eq.~\ref{disorder_tc} up to substitution $\Gamma_0 \rightarrow \bar{\Gamma}$ and $n \rightarrow 4$. We have shown that correspondence between critical temperature and robustness against the disorder is the general feature of the superconductivity.

Superconducting fitness function in case of the system with time-reversal symmetry is written as $F_c=[H_N,A]$, see Ref.~\cite{Ramires2018}  We rewrite fitness function $F_c=\sum F_{ci}$ in a spectral representation where $F_{ci}=\epsilon_i [P_i,A]$. We introduce partial fitness function $F_{pc}=\sum F_{ci}$ where sum $i \in {\bf p_l}$ is taken over the states that corresponds to band with the energy $\epsilon_i$ that crosses Fermi level. Following expression $\textrm{Tr}\langle F_{pc}^{\dagger}F_{pc}\rangle/\epsilon_i=1-\zeta$ establishes connection between superconducting fitness $F_c$ and parameter $\zeta$. If superconducting state is perfectly fit $F_c=0$ then parameter $\zeta=1$ that ensures robustness against the disorder~\cite{Andersen2020,Timmons2020}. This case can happen if Hamiltonian of the normal state commutes with the matrix structure of the order parameter $[H_N,A]=0$. S-wave order parameter $\hat{A}=\hat{1}$ always satisfy this condition that leads to the Anderson theorem~\cite{Anderson1959}. If matrix structure of the superconducting order parameter is the integral of motion then this superconductivity is also robust against the disorder~\cite{Michaelli2012}.

\begin{widetext}
\begin{table*}\label{table_zeta}
\begin{tabular}{|l|l|l|l|}
        \hline
        &  $\hat{\Delta}_2$ & $\hat{\Delta}_3$ & $\hat{\Delta}_4$
                \\ \hline
       \textrm{Representation} & 
$A_{1u}$  &   $A_{2u}$ & $E_u$ \\ \hline
         \textrm{Matrix structure} $\hat{A}$& $\sigma_ys_z$ &  $\sigma_z$ & $(\sigma_ys_x,\sigma_ys_y)$            \\ \hline
         $F_c/2i$& $R_1\sigma_z-m\sigma_xs_z$ & 
$v_zk_z\sigma_x-v(k_ys_x-k_xs_y)\sigma_y-R_1s_z\sigma_y$ & $(vk_y\sigma_z-m\sigma_xs_x,-vk_x\sigma_z-m\sigma_xs_y)$ \\ \hline
         $\zeta$& $\langle \frac{v_z^2k_z^2+v^2k_x^2+v^2k_y^2}{\mu^2} \rangle $&  $\langle \frac{m^2}{\mu^2} \rangle$ & $(\langle \frac{m^2+v^2k_x^2+R_1^2}{\mu^2} \rangle,\langle \frac{m^2+v^2k_y^2+R_1^2}{\mu^2} \rangle)$ \\ \hline
    \end{tabular}
             \caption{Superconducting fitness function $F_c$ and parameter $\zeta$ for possible odd-parity superconducting pairings taken from Ref.~\cite{Fu2010}. Here $\langle ... \rangle$ means average over the Fermi surface of the normal state.}
              \end{table*}
\end{widetext}
            
\begin{figure}[ht]
\includegraphics[width=1\linewidth]{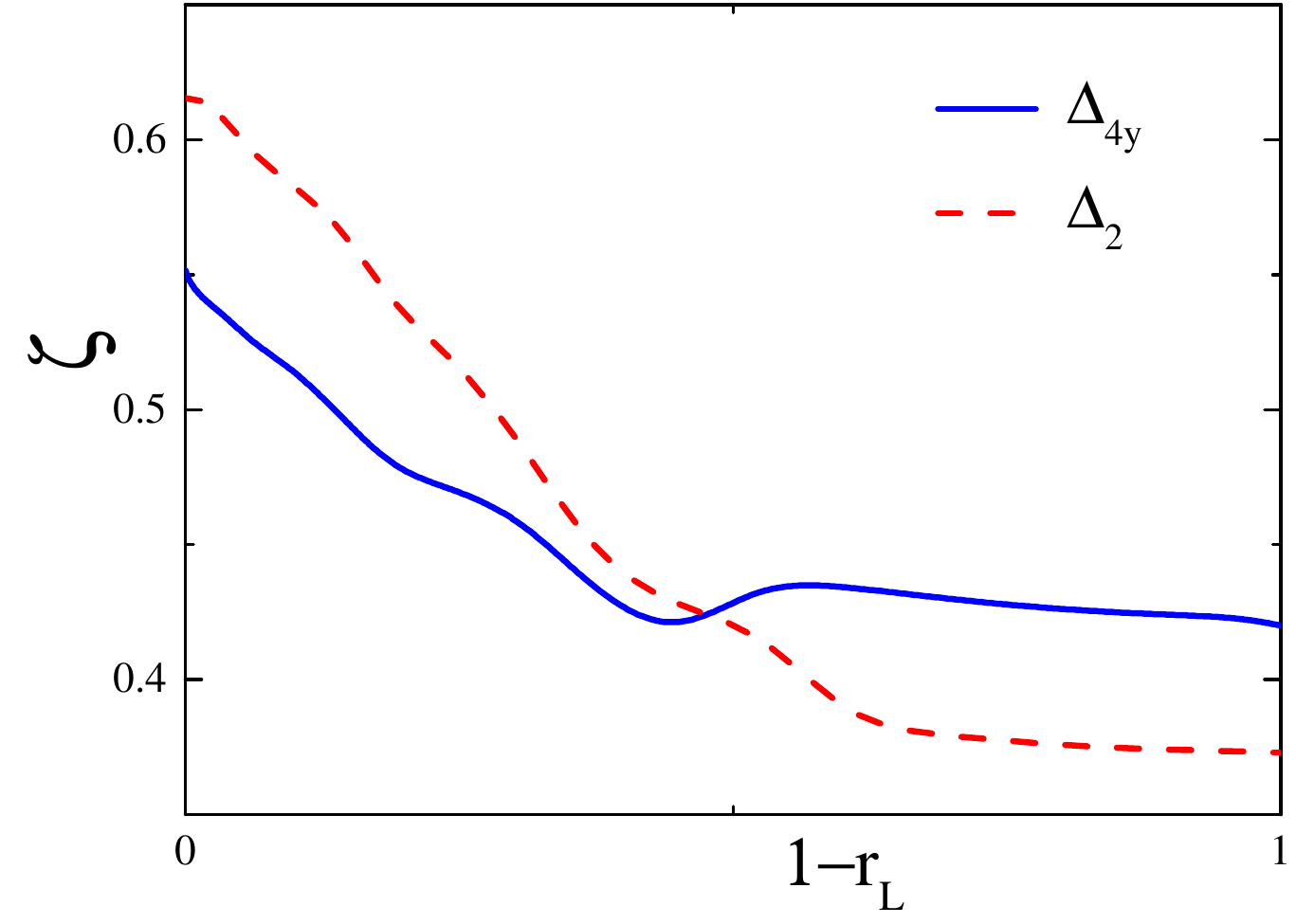}
\caption{Parameter $\zeta$ as a function of Lifshitz parameter $r_L$ for nematic state $\Delta_4$ and orbital triplet state $\Delta_2$.  }
\label{lifshitz_zetar}
\end{figure}

Parameter $\zeta$ is a useful quantity: it shows how the symmetry of the order parameter affects its critical temperature and its robustness against the disorder. We calculate parameter $\zeta$ along with the superconducting fitness function $F_c$ for different possible odd-parity superconducting order parameters for the topological insulator with the hexagonal warping $H_N+R_1 s_z\sigma_z$ where $R_1=\lambda k_x(k_x^2-3k_y^2)$. Results are summarized in Table~\ref{table_zeta}. Terms that contribute to the superconducting fitness $F_c$ decrease the critical temperature of the corresponding order parameter. We see that large single-electron gap $m$ disfavors $\Delta_2$ and nematic $\Delta_4$ order parameters and stimulates $\Delta_3$. Hexagonal warping stimulates nematic superconductivity\cite{Akzyanov2020} $\Delta_4$ that allows it to win against $\Delta_2$. This analysis is similar to one from Ref.~\cite{Ramires2018}. Parameter $\zeta$ as a function of the Lifshitz transition parameter $r_L$ for $\Delta_2$ and $\Delta_4$ order parameters is shown at Fig.~\ref{lifshitz_zetar}. We see that $\zeta$ decreases with the transformation of the Fermi surface from closed one to cylindrical for both states. At some point, the Lifshitz transition makes critical temperature for $\Delta_4$ higher than $\Delta_2$. This effect occurs due to an effective increase of warping for an open Fermi surface. Thus, we conclude that the Lifshitz transition helps the nematic state to compete against other odd-parity order parameters. However, the Lifshitz transition decrease parameter $\zeta$ for the nematic state while for the s-wave order parameter this quantity is unaffected. Thus, the open Fermi surface helps the s-wave order parameter the most.

\section{Meissner current}\label{sec_meissner}
In this section we show how disorder and shape of the Fermi surface affect Meissner current in the nematic superconductor. Superconducting current in a linear response is proportional to the vector potential and superconducting density $J_{\alpha} =-n_s A_{\alpha}$. We express current as $J_{\alpha}= -\sum_\beta K_{\alpha\beta}A_{\beta}$ where Meissner kernel is\cite{Schmidt2020}
\begin{equation}\label{meissner_kernel}
K_{\alpha\beta}=-T\sum\limits_{k,\omega} v_{\alpha} F v_{\beta} \bar{F}.
\end{equation}
Current operator $v_{\alpha}=-\partial H_N(k_{\alpha}-A_{\alpha})/\partial A_{\alpha}=\partial H_N(k_{\alpha})/\partial k_{\alpha}$ coincides with the velocity operator in case of linear spectrum, $\alpha=x,y,z$.
In general, we should use full Green's function $\hat{G}$ in Eq.~\ref{meissner_kernel} and then subtract the contribution of the normal part. In our case of linearized Green's functions, it means that we keep the anomalous part of Green's function only. As it is shown in Ref.~\cite{Schmidt2020} this procedure is correct even if we calculate response beyond linearized in $\hat{\Delta}$ theory.

\begin{figure}[ht]
\includegraphics[width=1\linewidth]{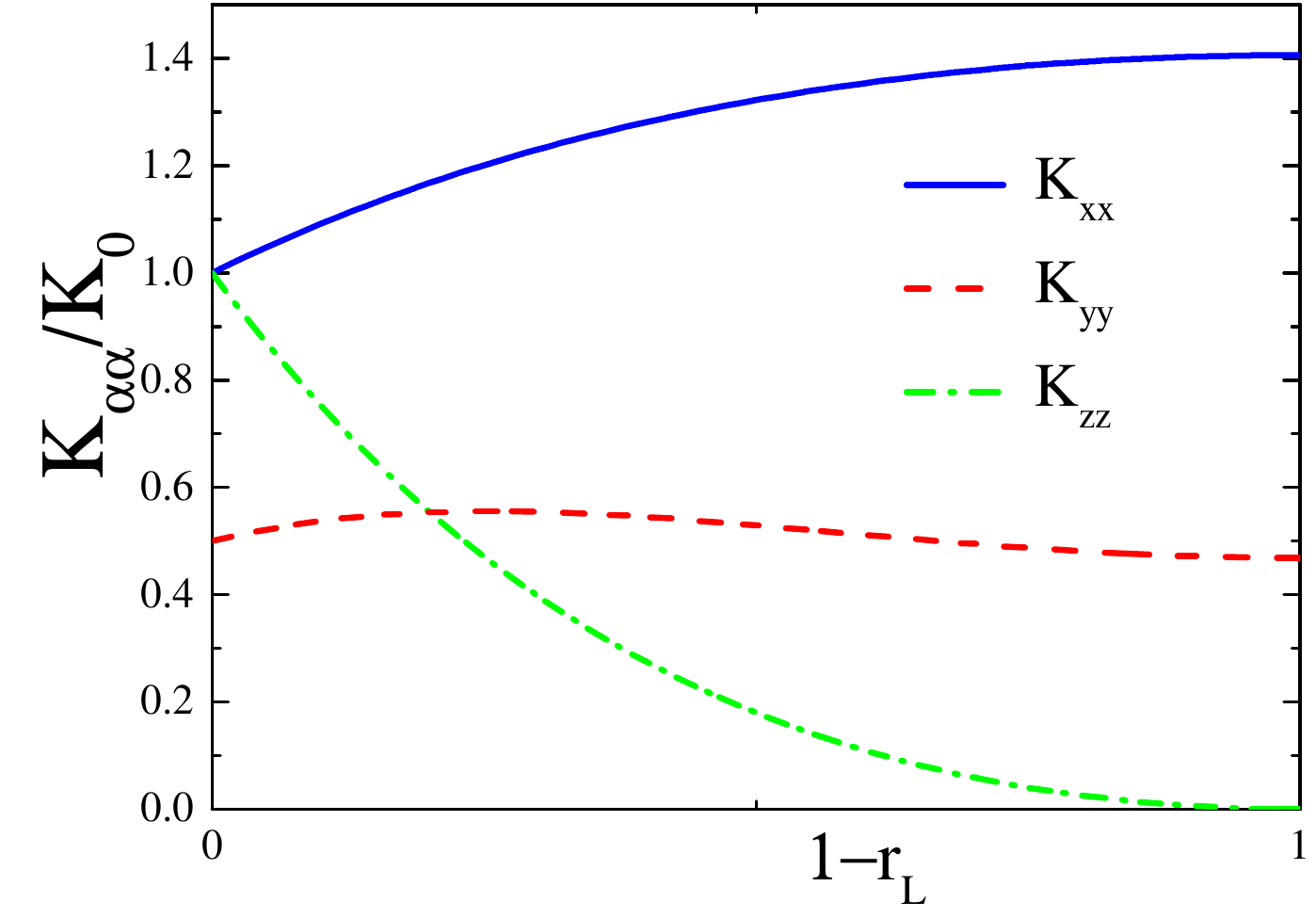}
\caption{Meissner kernels $K_{\alpha\alpha}$ as functions of Lifshitz parameter $r_L$ for $\Delta_x$ nematicity direction.}
\label{lifshitz_kernel}
\end{figure}

First, we compute correlation function in a clean limit near the critical temperature using Eq.~\ref{anomalous_green_function} for the anomalous Green's function. We consider only $\Delta_x$ orientation. Straightforward calculations gives us following expressions
\begin{eqnarray}
K_{xx}=\frac{45-10r_L^2-3r_L^4}{32}K_0, \nonumber\\
K_{yy}=\frac{15+10r_L^2-9r_L^4}{32}K_0, \nonumber\\
K_{zz}=\frac{v_z^2r_L^2(5+3r_L^2)}{8v^2}K_0.
\end{eqnarray}
\begin{equation}\label{corr_function}
K_0=\frac{v^2\pi\rho(\mu)(1-m^2/\mu^2)^2}{15}T\sum_{\omega} \Delta^2/|\omega|^3
\end{equation}
Integration over the Matsubara frequencies gives us $K_0=\kappa \Delta^2/T^2$ where $\kappa =v^2\rho(\mu)(1-m^2/\mu^2)^2 7\zeta(3)/120\pi^2 $. Here $\zeta(3)\simeq 1.2$ is the Riemann zeta function.
In case of $r_L=1$ for closed Fermi surface we get $K_{xx}=2K_{yy}=K_{zz}v_z^2/v^2=K_0(\omega)$ which is similar to the results of Ref.~\cite{Schmidt2020}. In case of $r_L=0$ for cylindrical Fermi surface we have $K_{xx}=3K_{yy}=3K_0(\omega)$ and $K_{zz}=0$. We plot correlation functions $K_{\alpha\alpha}$ as a function of the parameter of the Lifshitz transition $r_L$ in Fig.~\ref{lifshitz_kernel} for $v_z/v=2/3$. We see that Lifshitz transition increases anisotropy of the response.

\begin{figure}[ht]
\includegraphics[width=1\linewidth]{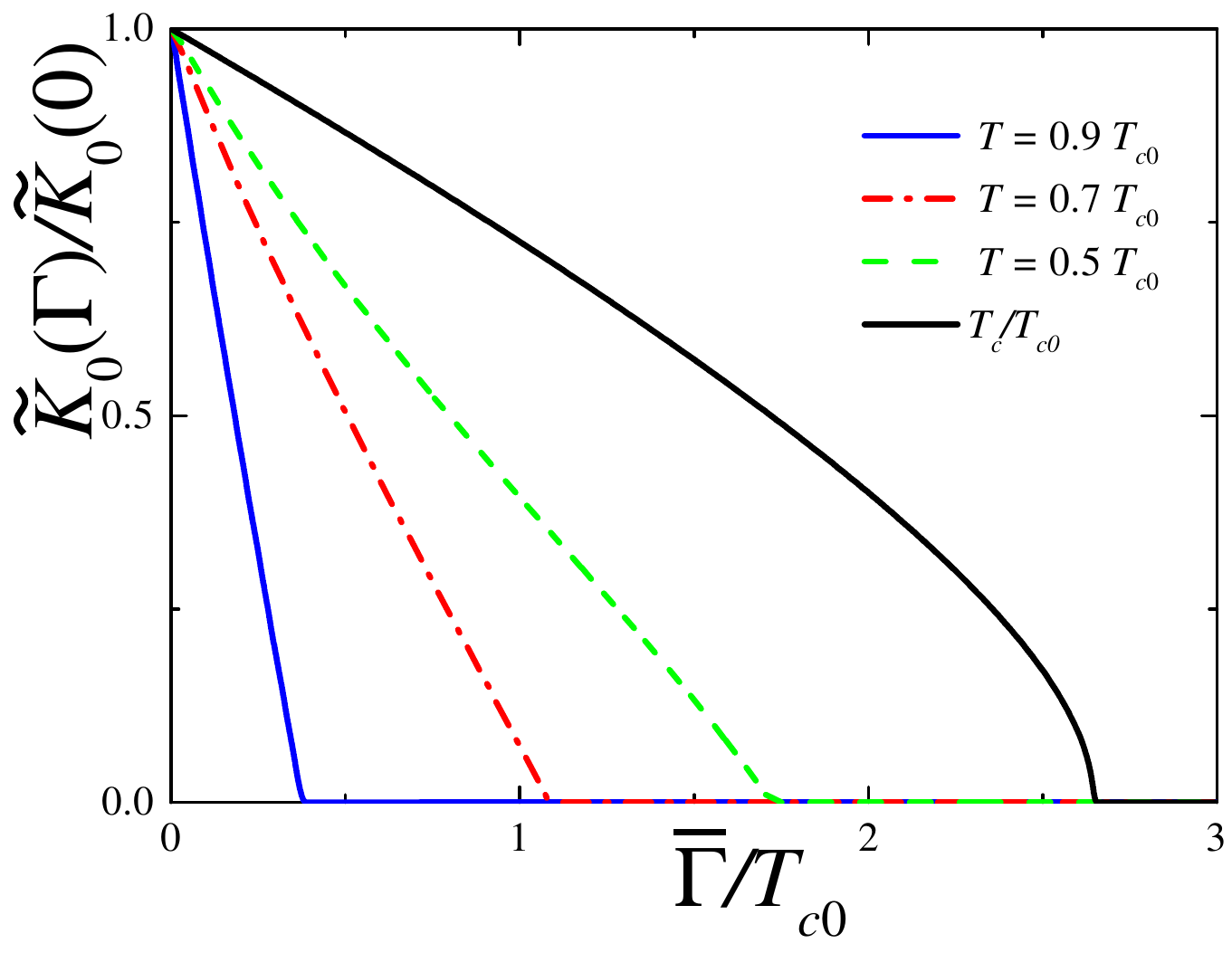}
\caption{Meissner kernel $\tilde{K_0}$ as function of disorder $\bar{\Gamma}/T_{c0}$ for different values of temperature $T$ for $\zeta=2/3$. Black line corresponds to the dependence of the critical temperature $T_c$ from the disorder $\bar{\Gamma}$. }
\label{lifshitz_meissnerg}
\end{figure}

We can introduce disorder by substitution $\omega \rightarrow \tilde{\omega}$ and $\Delta \rightarrow \tilde{\Delta}$, where $\tilde{\omega}$ and $\tilde{\Delta}$ are determined by Eq.~\ref{disorder_renormalized_delta}. Inserting this into Eq.~\ref{corr_function} gives us following expression for the Meissner kernel $\tilde{K}_0$ in disordered case
\begin{eqnarray}
\tilde{K}_0=\kappa\sum_\omega \frac{\Delta^2}{(\omega+\bar{\Gamma})(\omega+\bar{\Gamma}(1-\zeta))^2}.
\end{eqnarray}
We can see from this expression that superconducting density is suppressed by the disorder even if the critical temperature is robust $\zeta=1$. In case of $\bar{\Gamma} \ll T_c \ll \bar{\Gamma}/(1-\zeta) $ critical temperature is unaffected by the disorder while Meissner currents are suppressed $\tilde{K} \propto 1/\bar{\Gamma}$. In case of large disorder $T_c \ll \bar{\Gamma}/(1-\zeta) $ superconducting density is suppressed because of the suppression of the critical temperature by the disorder $\tilde{K} \propto \Delta^2 \propto T_c-T$. We plot the Meissner kernel as a function of disorder at Fig.~\ref{lifshitz_meissnerg}. We see that the superconducting density is suppressed stronger than the critical temperature. Anisotropy of the Meissner currents remains the same as in the clean case.

\section{Discussion}\label{sec_lifshitz_discussion}
In Ref.~\cite{Almoalem2021} the authors state that the superconductivity in doped topological insulators appears along with the Lifshitz transition from closed to open Fermi surface. Open Fermi surface has been observed in different compounds of doped topological insulators with the nematic superconductivity~\cite{Lahoud2013,Lawson2016,Kuntsevich2019a}. In our work, we get that the Lifshitz transition is destructive for the nematic superconductivity. Both critical temperature and robustness against the disorder are smaller for the open Fermi surface than for the closed one if the density of states is the same. This connection between the shape of the Fermi surface and superconducting properties can be one of the reasons why the critical temperature in doped topological insulators is insensitive to the carrier density~\cite{Kuntsevich2019a}.   

Nematic superconductivity only partially robust against the potential disorder and is suppressed if the disorder is large. Our results are consistent with the Refs.~\cite{Cavanagh2020,Sato2020,Dentelski2020} and are in disagreement with the Ref.~\cite{Andersen2020}.
We derive that critical temperature in clean case and robustness against the disorder are closely tied. The connection between these quantities comes from the mutual symmetry between the Hamiltonian of the normal state and the spin-orbital structure of the superconducting order parameter. We express this connection through superconducting order parameter that is projected onto the states of the Hamiltonian of the normal state, see Eq.~\ref{proj_order_parameter}. This quantity is closely tied with the conception of the superconducting fitness~\cite{Ramires2018}, see Sec.~\ref{lifshitz_sec_spectral}. A similar connection between robustness against the disorder and superconducting fitness has been derived for the s-wave states~\cite{Cavanagh2020}. In Refs.~\cite{Timmons2020,Andersen2020} connection between superconducting fitness and robustness against the disorder has been discussed as well.  

In Ref.~\cite{Kawai2020} Cu doped of Bi$_2$Se$_3$ samples show two-fold behavior of the second critical field that is a distinctive feature of the nematic superconductivity~\cite{Kawai2020}. At large doping carrier density substantially increases and two-fold symmetry of the second critical field $H_{c2}$ disappears. It was suggested that this occurs due to phase transition to the different superconducting states. An increase of the chemical potential gradually transforms the Fermi surface into the cylindrical one. This process makes the nematic superconducting state less favorable in comparison with the even-parity s-wave superconducting state. Thus, we conclude that the most likely superconducting state in Cu overdoped Bi$_2$Se$_3$ without two-fold symmetry of $H_{c2}$ is even parity s-wave.

We found that the Meissner current near the critical temperature is diamagnetic and anisotropic that confirms the results of Ref.~\cite{Schmidt2020}. Meissner current is largest along the nematicity direction for closed Fermi surface. This anisotropy is increased by the Lifshitz transition $K_{xx}=3K_{yy}$ for $r_L=0$. In general, anisotropic superconductors have quite complex behavior in a magnetic field~\cite{kopnin_book}. We assume a simplified situation that London penetration length for the magnetic field applied along $i$ direction expresses through Meissner kernel $\lambda_{i}^2 \propto 1/K_{ii}$ as for the isotropic superconductor. We are not aware of the works on the in-plane anisotropy of the first critical field, so we focus on the anisotropy between averaged in-plane $\lambda_{ab}^2=(\lambda_x^2+\lambda_y^2)/2$ and out-of-plane penetration lengths $\lambda_c^2=\lambda_z^2$ that is $\kappa=\lambda_c^2/\lambda_{ab}^2$.  We take $v_z/v=2/3$ from Ref.~\cite{Liu2010}. In case of closed Fermi surface our calculations lead to $\kappa=1/2$ for $v_z/v=2/3$. In Ref.~\cite{Matano2016} this anisotropy parameter $\kappa\simeq2.4$ while in Ref.~\cite{Fang2020} $\kappa\simeq2.6$. As we can see, the assumption of a closed Fermi surface is inconsistent with the experimental results. We found that $r_L \sim 1/2$ gives experimentally relevant anisotropy of the first critical field. In this case, Fermi surface is a corrugated cylinder.

In Ref.~\cite{Kriener2012} it was obtained that superconducting density is suppressed by the disorder while the critical temperature is largely unaffected. In Ref.~\cite{Smylie2017} authors conclude that only a small part of scattering events contribute to the depairing of Cooper's pairs. This situation occurs since critical temperature $T_c$ is suppressed by effective disorder $(1-\zeta)\bar{\Gamma}$ and only $1-\zeta$ of scattering events contribute to the depairing. However, every scattering event contributes to the suppression of the superconducting density similar to the case of s-wave superconductor~\cite{levitov_book}. Thus, superconducting density is suppressed stronger by the disorder than the critical temperature.

In general, strong Coulomb repulsion that is accompanied by the fluctuations in $E_u$ channel can lead to the mixing between singlet s-wave order parameter and triplet nematic order parameter within $E_u$ representation~\cite{Wu2017a}. However, due to large dielectric constant Coulomb repulsion is weak in topological insulators. Note, that presence of the fluctuations $E_u$ only does not lead to such coupling~\cite{Kozii2015}.  

In conclusion, we get that Lifshitz transition from closed to open Fermi surface affects both critical temperature and robustness against the disorder in nematic superconductors. We found that critical temperature in a clean limit and robustness against the disorder are tied through the superconducting fitness. Anisotropy of Meissner currents is increased by the Lifshitz transition.

\section*{Acknowledgment}
We acknowledge support by the Russian Scientific Foundation under Grant No 20-72-00030 and partial support from the Foundation for the Advancement of Theoretical Physics and Mathematics “BASIS”. 

\bibliography{disordered_nematic}
\end{document}